# The Sustainable Response Strategy to COVID-19: Pandemic Urban Zoning Based on Multimodal Transport Data

**Yufei Wang, Research Assistant**
Jiangsu Key Laboratory of Urban ITS, Southeast University
Jiangsu Province Collaborative Innovation Center of Modern Urban Traffic Technologies, Southeast University
School of Transportation, Southeast University, Nanjing, China, 211189
Email: wangyufei@seu.edu.cn

**Mingzhuang Hua, Ph.D. Candidate**
Jiangsu Key Laboratory of Urban ITS, Southeast University
Jiangsu Province Collaborative Innovation Center of Modern Urban Traffic Technologies, Southeast University
School of Transportation, Southeast University, Nanjing, China, 211189
Email: huamingzhuang@seu.edu.cn

**Xuewu Chen (Corresponding Author), Ph.D., Professor**
Jiangsu Key Laboratory of Urban ITS, Southeast University
Jiangsu Province Collaborative Innovation Center of Modern Urban Traffic Technologies, Southeast University
School of Transportation, Southeast University, Nanjing, China, 211189
Email: chenxuewu@seu.edu.cn

**Wendong Chen, Ph.D. Candidate**
Jiangsu Key Laboratory of Urban ITS, Southeast University
Jiangsu Province Collaborative Innovation Center of Modern Urban Traffic Technologies, Southeast University
School of Transportation, Southeast University, Nanjing, China, 211189
Email: chenwendong@seu.edu.cn

**Long Cheng, Ph.D., Associate Professor**
Jiangsu Key Laboratory of Urban ITS, Southeast University
Jiangsu Province Collaborative Innovation Center of Modern Urban Traffic Technologies, Southeast University
School of Transportation, Southeast University, Nanjing, China, 211189
Email: longcheng@seu.edu.cn

Word Count: 5830 words + 5 tables = 7080 words

*Submitted [29[th] July, 2022]*

*Wang, Hua, Chen, Chen and Cheng*

**ABSTRACT**

Since the outbreak of COVID-19, it has rapidly evolved into a sudden and major public health emergency globally. With the variants of COVID-19, the difficulty of pandemic control continues to increase, which has brought significant costs to the society. The existing pandemic control zoning method ignores the impact on residents' lives. In this study, we propose a refined and low-cost pandemic control method by scientifically delineating zoning areas. First, a spatial interaction network is built up based on the multimodal transport travel data in Nanjing, China, and an improved Leiden community detection method based on the gravity model is used to obtain a preliminary zoning scheme. Then, we use spatial constraints to correct the results with the discrete spatial distribution. Finally, reasonable zones for pandemic control are obtained. The modularity of the algorithm results is 0.4185, proving that the proposed method is suitable for pandemic control zoning. The proposed method is also demonstrated to be able to minimize traffic flows between pandemic control areas and only 24.8% of travel connections are cut off, thus reducing the impact of pandemic control on residents' daily life and reducing the cost of pandemic control. The findings can help to inform sustainable strategies and suggestions for the pandemic control.

## 1. INTRODUCTION

Since the outbreak of COVID-19 in December 2019, due to its rapid spread and strong infectivity, it has rapidly evolved into a sudden and major public health emergency globally (*1*). As of June 8, 2022, the cumulative number of confirmed COVID-19 cases worldwide reached 533,047,318 and the number of deaths reached 6302,274 (*2*). The normality of the pandemic has become the current global trend. The world has also entered the post-pandemic era, and the fight against the virus continues. However, the continued spread of the virus has led to increasing difficulties in the control of COVID-19. At the same time, frequent population movements make it difficult to implement the control of COVID-19, and the cost of anti-pandemic continues to rise. Under the influence of the pandemic, the GDP of various countries in the world has been affected to varying degrees. Due to COVID-19 pandemic, numerous people lose their jobs, and people's living standards decline. All of these factors lead to a change in consumers' attitudes and confidence, further resulting in a depression in the national economy (*3*). To this end, CDC proposed COVID-19 sustainable response planning to deal with the COVID-19 pandemic in the worldwide (*4*). Therefore, in the context of the normality of COVID-19, reducing the economic losses of the whole society due to the pandemic, and then achieving sustainable anti-pandemic policies has become an urgent problem that we need to solve.

With the concentrated outbreak of COVID-19, many cities in China have been forced to adopt closed-off management to block the spread of the coronavirus. Closed-off management is divided into full closure of the city and zoning closed of the city. Since January 24, 2020, seven cities in Hubei have been closed. After that, Lanzhou, Guangzhou, Zhengzhou, and other cities have experienced full closure. However, the full closure of the city against COVID-19 will greatly limit the economic development of the city and seriously affect the daily life of the residents. And the difficulty of COVID-19 control is closely related to the size of the population and area of the control region. Therefore, against the background of the normality of the pandemic, the original method to fight COVID-19 brings intensive barriers to the lives of residents. Pandemic control zoning refers to a policy that relies on mobility restrictions and public health measures that are solely based on the epidemiological status of well-identified zones (*4*). Pandemic control zoning aims at reducing the spread of infectious disease and, at the same time, allowing zones where the virus is under control to levy restrictions and return to normal activity (*5*). In March 2022, COVID-19 broke out in Shanghai, China, and Shanghai implemented zoning control measures on March 28. Shanghai is divided into two areas, Pudong and Puxi, according to the geographical boundary of the Huangpu River. The two areas are screened for all employees, one area is restricted except that travel for nucleic acid testing is permitted, and the other area is normal for work and life. Shanghai has adopted the method of zoning control, which is a very creative pandemic control measure.

Regarding zoning control, many cities have adopted zoning control methods based on geographic separation or administrative borders (*6,7*). But the pandemic control zoning plan based on administrative boundaries often incur high societal costs. Administrative regions are from historical evolutions and cannot represent the movement pattern of urban residents (*8*). The delineation of administrative borders does not take transportation connections into account, so relying on administrative borders for pandemic control zoning are not very cost-effective. Therefore, it is necessary to use transport big data analysis to develop more scientific and reasonable zoning plans. As such, full closure of the city can be avoided and refined pandemic control solutions would be obtained.

To handle on these issues, this paper proposes a pandemic zoning method based on Nanjing's multimodal travel data to minimize percentage of cut-off trips between partitions. The paper contributes to the literature in three ways. First, this paper uses community detection algorithm combined with big data analysis for scientific pandemic control zoning strategy. Second, this paper quantitatively analyzes the impact of the COVID-19 pandemic on residents' travel. Third, the method proposed in this paper helps to reduce the cost of pandemic control and achieve sustainable response to the COVID-19 in the post-pandemic era.

This paper is organized as follows. Section 2 introduces the related works of COVID-19 pandemic control and community detection algorithm. Section 3 explains data sources and data processing method. Section 4 displays the methodology for pandemic control zoning. Section 5 presents analyses results. Finally, Section 6 concludes and discusses this study.

## 2. LITERATURE REVIEW

In the context of COVID-19, pandemic control policies have attracted the attention of many scholars in recent years. Yang et al. (9) found that maintaining social distance between susceptible and infected populations can effectively reduce the number of infected cases. Through real-time epidemic surveillance data from Baidu, they also found that population migration and frequent population movement are positively correlated with the spread of COVID-19. Hua et al. (10) re-emphasized the need to maintain the necessary social distance in public transportation during the COVID-19 pandemic to help reduce the risk of infection. Some studies have shown that large-scale nucleic acid testing, the establishment of modular hospitals, and strict control measures can effectively control the pandemic. This method can avoid the large-scale spread of COVID-19, and reduce the number of infected cases and deaths (*11*). Velthuis and Bajardi et al. (*12,13*) pointed out that movement restriction can reduce the probability and frequency of the spread of the pandemic and is an effective means of controlling the pandemic. Some studies have further pointed out that the quarantine of cities can effectively reduce the peak number of COVID-19 infections and delay the peak time of infections (*14*). The implementation of large-scale pandemic control at the city level in Wuhan, China effectively curbed the spread of the COVID-19 (*15*). However, when the number of confirmed cases increases exponentially, large-scale pandemic control measures at the city level make it difficult to fully trace all contacts for isolation and epidemiological investigation (*16,17*). Rubin et al. (*18*) pointed out that widespread containment may frighten residents and constrain many individuals with no significant risk of diseases. In addition, large-scale pandemic containment requires enormous resources, including medical and security services, to maintain the lives of people in containment areas (*19*). Thus, zoning control has been preliminarily discussed. By implementing zoning closed management and gradual screening of residential communities, the source of infection can be effectively controlled and the transmission route of the virus can be cut off. After 14 days of closed management of residential community, the number of close contacts and people under medical observation decreased. The results show that the measure saves a lot of medical and social resources (*20*). Lipsitch et al. (*21*) found that formulating pandemic control standards and policies according to the development of the epidemic situation in different regions can help save medical and financial resources. Kuo et al. (*22*) achieved regionalization for infection control through a containment zone delineation algorithm considering the regularity of human mobility. It significantly delays the pandemic peak and decreases the severity of the pandemic. However, Kuo (*22*) adopted the demand method based on the resident survey data. Hua et al. (*23*) pointed out that the trip records data is superior to the user survey data in terms of the more accurate resource of travel behavior. Therefore, the pandemic control zoning research based on multimodal traffic travel data is a promising research direction.

The zoning of the pandemic control area belongs to the research category of community detection. Community detection is often used to analyze the structure of large complex networks. One of the most popular algorithms for uncovering community structure is the Louvain algorithm (*24*). But the Louvain algorithm may yield arbitrarily badly connected communities. In the worst case, communities may even be disconnected, especially when running the algorithm iteratively. To solve this problem, the Leiden algorithm improves it. The Leiden algorithm is faster than the Louvain algorithm and uncovers better partitions (*25*). Some scholars have improved the above algorithms based on spatial constraints. Adding geospatial constraints can make the community detection results more closely match the actual spatial-geographic relationships (*26,27*). Based on the above algorithms, many scholars have conducted network analysis research and used it for pandemic control research. Dai et al. (*28*) used community detection algorithms to study population migration patterns and spatio-temporal evolution mechanism under the background of COVID-19. Wickramasinghe et al. (*29*) applied the network analysis method to explain the spread of the COVID-19 virus based on data from the first three months of the 2020 COVID-19 outbreak across the world and in Canada. Community detection algorithms are utilized to understand the spread of COVID-19 among countries and the impact of other countries on the spread of COVID-19 in Canada. Social networks are also being used to study the spread of the epidemic through community detection algorithms. Azad et al. (*30*) studied how the COVID-19 spread across Indian states through social groups. However, few of the existing studies analyzed pandemic from the perspective of community detection. And in these few studies, the main focus is on the spread of COVID-19, but few use community detection algorithm to study pandemic control. Besides, in existing research, traffic big data is mainly used to study the spread of the virus and changes in travel behavior before and after the pandemic, and is rarely used in pandemic response research. Using residential travel data to study the zoning of pandemic control areas through community detection method remains a critical research gap. Thus, changing the pandemic control method that relies on experience through partitioning the pandemic control area scientifically based on travel big data analysis is the research purpose of this paper, which help to inform refined, low-cost, and sustainable pandemic control measures.

In the post-pandemic era, it is of great significance to study the network structure and spatio-temporal evolution mechanism of residents' travel by complex network analysis. However, existing research on pandemic control zoning lacks quantitative studies on the impact of a pandemic on travel by residents. Partitioning of pandemic control areas through community detection methods based on multimodal urban transport data can maximize the satisfaction of residents' travel demands within the pandemic control areas and minimize the cut of travel connections between pandemic control areas. Such pandemic control areas can reduce the impact of COVID-19 on residents' daily life, and control the spread of COVID-19 effectively. In this study, a spatial interaction network is constructed from multimodal transport travel data, and the spatially constrained Leiden algorithm is utilized to identify pandemic control areas. This paper is the first to study urban pandemic control zoning methods based on multimodal transport data, which provides references and suggestions for low-cost and sustainable control of COVID-19 in the post-pandemic era.

## 3. DATA
### 3.1 Study Area
Nanjing is in the east of China, the capital of Jiangsu Province, an important national research and education base and comprehensive transportation hub. Nanjing has 11 districts with a total area of 6,587.02 square kilometers and a resident population of 9,423,400 by the end of 2021. As one of the first batch of Transit Metropolis launched in China, Nanjing has a well-developed multimodal transport system. Nanjing Metro has 11 lines and 191 stations, with a total line length of 427.1 kilometers, forming a metro network covering the whole city. The administrative districts Gaochun and Lishui are far away from the main urban area, so in this paper, the study area is the 9 main districts in Nanjing, as shown in Figure 1.

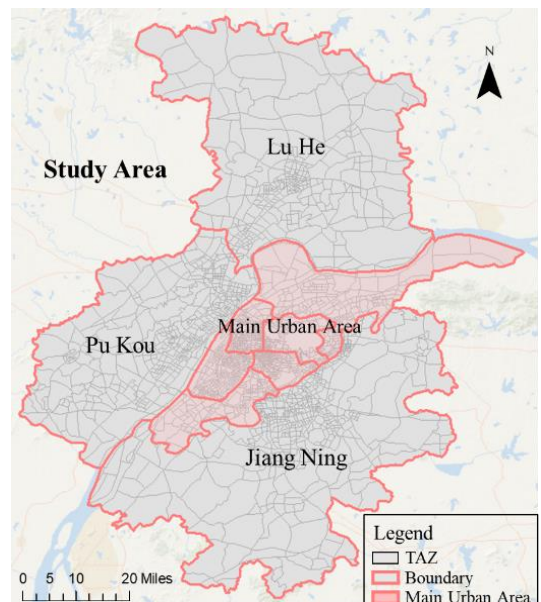

**Figure 1 Spatial distribution of Nanjing study area**

### 3.2 Data Source
Two types of datasets are adopted in this paper. The first type of data is the multimodal transport travel data of Nanjing residents, and the other is the information about traffic analysis zones (TAZ). The first category of data includes four types of travel data: free-floating bike-sharing (FFBS) journey data, Nanjing public bike data, Nanjing metro travel data, Nanjing smart card data from bus. These four types of data include the following fields: ID, date, origin location, destination location, origin time, destination time, as shown in Table 1 (a). The second category of the dataset is traffic analysis zones (TAZ) data of Nanjing, including ID, population, area, employment, and geometry, as shown in Table 1 (b). TAZ information collected by the Baidu Huiyan platform derives from Baidu Map, one of the largest map service providers in China. A total of 3028 TAZs are obtained in the study area.

The data of 3,647,984 FFBS trips in Nanjing for one week (from November 10 to November 16 in 2020) are provided by the Nanjing bike-sharing supervision platform. 459,609 public bike trips in Nanjing for the same week are provided by Nanjing public bicycle company. 8,614,830 metro trips in Nanjing for one week (from March 10 to March 16 in 2019) are provided by Nanjing Metro Company. Since the average daily travel volume of Nanjing Metro between 2019 and 2020 is relatively stable, which is suitable for combined analysis. The smart card data of 6,514,543 trips from the bus service in Nanjing for one week (from July 10 to July 16 in 2020) are provided by Nanjing Bus Corporation. The travel datasets of multimodal transport modes contain a total of 16,701,097 trips. Some of the fields in the dataset are shown in Table 1.

During that period, local COVID-19 cases and imported cases continued to appear. Except for Liuhe Pukou and Gaochun, the rest of the administrative regions were once set as medium or high-risk areas. The pandemic situation was severe (*31*).

**Table 1 Dataset**
**(a) Public transport trips**

| Type | ID | Date | Origin | | Destination | |
|---|---|---|---|---|---|---|
| | | | Time | Location | Time | Location |
| FFBS | Mobike864**** | 2020-11-10 | 20:33:23 | (118.88459,32.08721) | 20:43:20 | (118.87801,32.08519) |
| FFBS | DiDibike115**** | 2020-11-12 | 11:11:28 | (118.77237,32.09874) | 11:20:16 | (118.75962,32.10166) |
| Public Bike | 1017**** | 2020-11-15 | 11:40:56 | (118.74804,32.03575) | 11:52:12 | (118.74847,32.04295) |
| Public Bike | 1073**** | 2020-11-10 | 15:20:18 | (118.72105,32.00288) | 15:26:58 | (118.72376,32.00612) |
| Metro | 0D401B**** | 2019-3-10 | 16:41:01 | (118.72796,31.98994) | 15:27:37 | (118.75692,31.99289) |
| Metro | 084FBF**** | 2019-3-11 | 08:36:14 | (118.79290,32.05934) | 08:49:27 | (118.73489,32.23092) |
| Bus | 1700758**** | 2020-7-10 | 09:19:12 | (118.75365,31.95445) | 09:31:24 | (118.75655,31.99061) |
| Bus | 1700758**** | 2020-7-16 | 21:10:56 | (119.06159,32.05293) | 21:27:14 | (118.86544,32.04216) |
| … | … | … | … | … | … | … |

**(b) TAZ information**

| TAZ ID | Area(m2) | Population | Employment | Geometry |
|---|---|---|---|---|
| 1 | 143597 | 5162 | 836 | Polygon (118.725801, 32.05027; …) |
| 2 | 74344 | 344 | 167 | Polygon (118.72942, 32.050106; …) |
| … | … | … | … | … |

**3.3 Data Processing**

The processing of the data includes the matching and conversion of the latitude and longitude of the travel origin and destination. The bus trip data does not include the drop-off location, so it is necessary to use the minimum entropy rate-improved trip-chain method for origin-destination estimation (*32*). Data cleaning mainly includes deduplication of data, deletion of data with null values, and deletion of anomalous data. Anomalous data mainly include the longitude and latitude drift, the abnormal travel distance, and the abnormal travel time.

To create multimodal transport spatial interaction networks, it is necessary to aggregate users' journeys into travel flows at the TAZ level. Through the spatial connection function of ArcGIS software, users' travel origins and destinations are matched to the corresponding traffic analysis zone to generate a joint data set, as shown in Table 2. It records the trip-related information from one TAZ to another TAZ. In addition to the travel flow, this dataset also records population, employment, and distance between any pair of TAZs. It is noteworthy that the distance used refers to the shortest network distance from the centroid of one TAZ to the centroid of another TAZ.

**Table 2 Travel flow dataset**

| Origin | | | Destination | | | Distance (km) | Flow (trips/week) |
|---|---|---|---|---|---|---|---|
| TAZ ID | Population | Location | TAZ ID | Population | Location | | |
| 1 | 5162 | (118.72580,3205027) | 2 | 344 | (118.72942,32.05011) | 1.3 | 14 |
| 1 | 5162 | (118.72580,3205027) | 10 | 5688 | (118.72393,32.04699) | 5.6 | 170 |
| … | … | … | … | … | … | … | .. |

## 4. METHODS

### 4.1 Spatial interaction network

To partition urban pandemic control areas, a spatial interactive network with multi-modal public transport travel data is developed in the first step. This network can be defined as a weighted graph *G (V, E, W)*, where *V* represents the set of all nodes in the network, and in this spatial interaction network is the centroid of all traffic analysis zones; *E* represents the set of edges between nodes in the network. In this spatial interaction network, it is the multimodal transport travel between all traffic analysis zones. The parameter *W* is the weight of the network edges. In this spatial interaction network, it is the sum of multimodal transport travel flow between all traffic analysis zones.

### 4.2 Improved Leiden algorithm

The spatial interaction network is analyzed by community detection algorithms. The mainstream community detection algorithm is a graph clustering method based on modularity optimization. Newman and Girvan developed the so-called modularity, which is one of the best-known quality measures for community detection (*33*). Its formula is as follows:

$$Q = \frac{1}{2m} \sum_{i,j} \left[ A_{ij} - \frac{k_i k_j}{2m} \right] \delta(C_i, C_j) \tag{1}$$

where $A_{ij}$ represents the edge weight between nodes *i* and *j*. $k_i$ is the sum of weights of edges connected to node *i*. Similarly, $k_j$ is the sum of weights of edges connected to node *j*; m represents the number of edges; $C_i$ is the community to which node *i* has been assigned, and $\delta(C_i, C_j)$ equals 1 if $C_i = C_j$, and 0 otherwise. Therefore, the value range of modularity is (0,1). Generally speaking, a higher modularity value will correspond to a steadier community structure. Good community detection results are expressed in the form of higher similarity of nodes within the community and lower similarity of nodes outside the community.

Louvain is a community detection algorithm with high sensitivity and computational speed based on modularity optimization. This algorithm is widely used in complex network structure analysis. But the Louvain algorithm may yield arbitrarily badly connected communities. In the worst case, communities may even be disconnected (*24*). Compared to the Louvain algorithm, the Leiden algorithm is faster than the Louvain algorithm and uncovers better partitions (*25*).

The Leiden algorithm consists of three phases: (1) local moving of nodes, (2) refinement of the partition (3) aggregation of the network based on the refined partition, using the non-refined partition to create an initial partition for the aggregate network. Leiden algorithm adds a Refine step to improve the partition results and prevent bad connections within the community. Figure 2 provides an illustration of this algorithm (*25*).

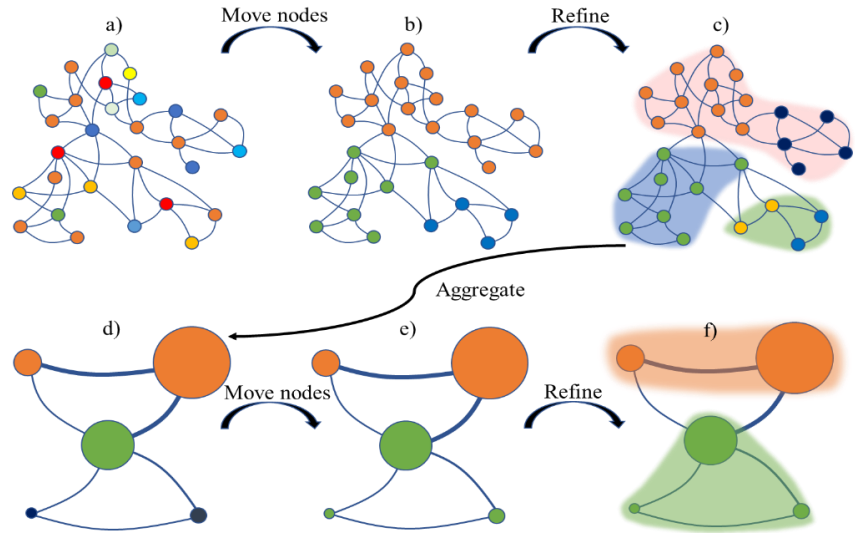

**Figure 2 The illustration of the Leiden algorithm (Source: Traag et al. (*25*))**

This paper proposes an improved Leiden algorithm based on the gravity model to detect communities with geospatial connections. Since geographic distance affects the strength of connections between nodes, geographic modularity is used to measure the partition performance. This modularity adds a weight of distance to edges when calculating modularity. It ensures spatial connectivity by attenuating the effects of distance, limiting community disconnections, and avoiding the division of oversized communities. According to the effect of distance decay, the geographic modularity is defined as:

$$Q = \frac{1}{2m} \sum_{i,j} \left[ \frac{A_{ij}}{d_{ij}^{\alpha}} - \frac{w_i w_j}{2m} \right] \delta\left(C_i, C_j\right) \tag{2}$$

$$w_i = \frac{k_i}{d_i}, w_j = \frac{k_j}{d_j} \tag{3}$$

where $A_{ij}$ represents the edge weight between nodes $i$ and $j$. $k_i$ is the sum of weights of edges connected to node $i$. Similarly, $k_j$ is the sum of weights of edges connected to node $j$. m represents the number of edges. $d_{ij}$ is the distance between node $i$ and node $j$. $d_i$ is the sum of distances of edges connected to node *i*. $d_j$ is the sum of distances of edges connected to node *j*. $C_i$ is the community to which node *i* has been assigned, and $\delta\left(C_i, C_j\right)$ equals 1 if $C_i = C_j$, and 0 otherwise. $\alpha$ is the decay coefficient. In this paper, we set $\alpha$ as 1.

In this model, the short-distance edge effect is strengthened, and the long-distance but less traveled edge is weakened by introducing geographical distance. It can ensure the spatial connectivity of the zoning control area results and avoid the emergence of mega-communities, by attenuating the distance effect. In this way, we consider both geographical distance and spatial connectivity as well as network modularity.

However, there is a strong connection of metro travel in the network between origin and destination. In addition, most of them are long-distance travel, which leads to spatially disconnected communities in the partition results. To further improve the community detection results and achieve the need for zoning pandemic control areas, we enforced spatial continuity and minimum area size rules to split spatially discontinuous communities. The core idea of the way to improve the algorithm is as follows, mainly adding three constraints to the Leiden algorithm.
(1) For an enclosed node with one neighboring node, merge it directly with the neighbor.
(2) For an enclaved node that has more than one neighboring node with connecting edges, merge it with the one that can achieve maximal positive modularity gain.
(3) For an orphan node without any edges to its neighboring nodes, group it with the one that has the smallest areal size (*27*).

## 5. RESULTS
In this paper, the spatial interaction network of multimodal transport trips in Nanjing is divided by the improved Leiden algorithm, which divides the city's 3,028 traffic analysis zones into 15 pandemic control areas. The function of this algorithm is to maximize the connections within the community and minimize the connections between the communities. The modularity of the algorithm results is 0.4185, indicating that it is reasonable to apply this algorithm to pandemic control zoning based on multimodal urban transport data, and the connection between partitions is relatively small. There is a high volume of travel between the traffic analysis zones where a metro station is located. This phenomenon leads to the edge weight of the spatial interaction network of the corresponding traffic analysis zones being relatively high. This results in the phenomenon of spatial disconnection in some communities, which does not meet the requirements for the zoning of pandemic control areas. The partition results are shown in Figure 3 (a). Therefore, we use the Leiden algorithm based on spatial constraints to correct the results with the discrete spatial distribution. The modularity of the algorithm results is slightly reduced to 0.3819. The final result of the pandemic control partition based on the community detection algorithm is shown in Figure 3 (b).

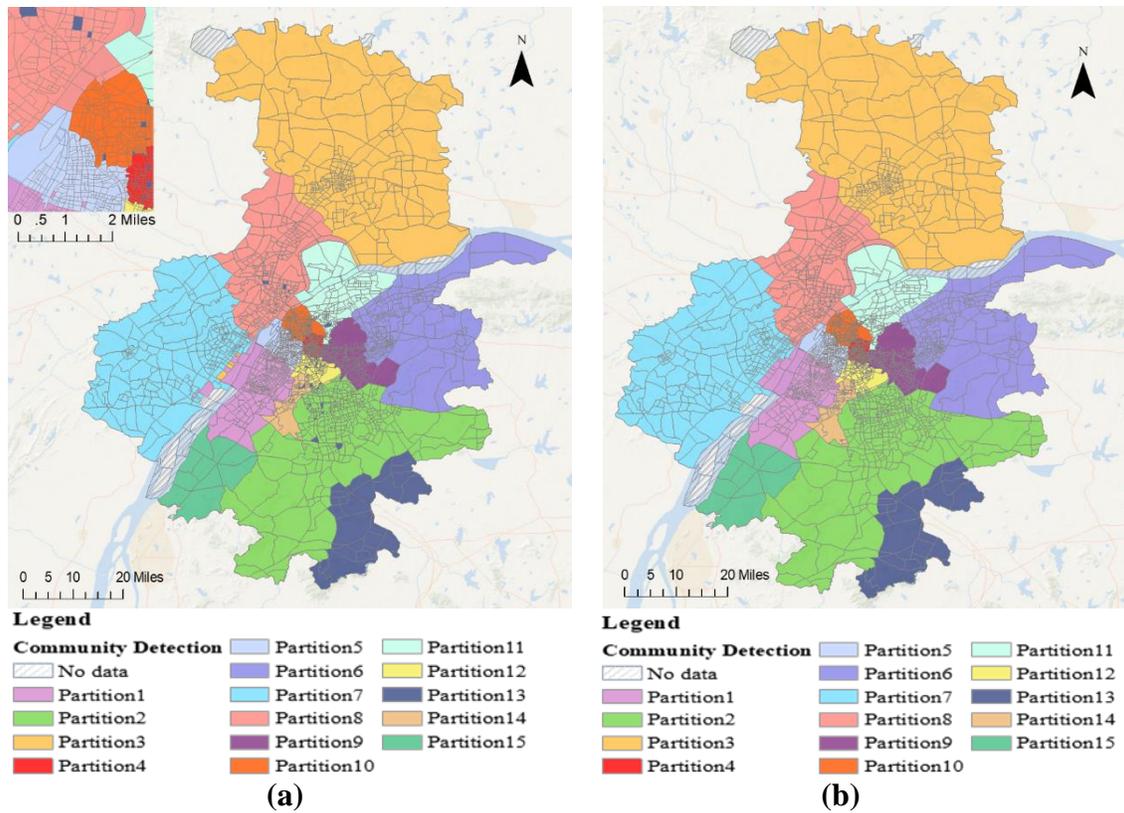

**(a)**　　　　　　　　　　　　　**(b)**

**Figure 3 (a) Results of zoning pandemic control areas by improved Leiden algorithm based on the gravity model and (b) corrected zoning result by spatial constraints**

The statistics of intra-partition trips and total trips are shown in Table 3. The population in partition 2 is the largest at 1,856,000. It can be seen that only 49.3% of travel connections are cut off through community detection and the average percentage of cut-off trips is 45.3%, which can effectively reduce the impact of pandemic control on residents' travel. And Table 4 shows the statistics of intra-administrative districts trips and percentage of cut-off trips. Compared with the zoning results proposed in this paper, pandemic control areas based on administrative districts cut off more travel connections between regions. And the variety of urban zoning is smaller, which is not conducive to more flexible pandemic response.

**Table 3 Statistics of travel within the pandemic control area**

| ID | Travel within the partition | Total trips in partition | Percentage of cut-off trips | ID | Travel within the partition | Total trips in partition | Percentage of cut-off trips |
|----|------|---------|-------|-----|---------|---------|-------|
| 1 | 814001 | 1445322 | 43.7% | 2 | 793516 | 1565796 | 49.3% |
| 3 | 203003 | 323750 | 37.3% | 4 | 1216952 | 2765344 | 56.0% |
| 5 | 626018 | 1409669 | 55.6% | 6 | 660339 | 1221855 | 46.0% |
| 7 | 181320 | 364057 | 50.2% | 8 | 786739 | 1453103 | 45.9% |
| 9 | 712767 | 1119151 | 36.3% | 10 | 965260 | 1561400 | 38.2% |
| 11 | 907676 | 1674548 | 45.8% | 12 | 830472 | 1111477 | 25.3% |
| 13 | 36331 | 63556 | 42.8% | 14 | 339629 | 510470 | 33.5% |
| 15 | 65767 | 111599 | 41.1% | Sum | 9139790 | 16701097 | 45.3% |

**Table 4 Statistics of travel within administrative districts**

| Name of Administrative district | Travel within the administrative district | Total trips in administrative district | Percentage of cut-off trips |
|---|---|---|---|
| JianYe | 1027089 | 1965139 | 47.7% |
| QinHuai | 1136900 | 2742648 | 58.5% |
| GuLow | 1268809 | 2418767 | 47.5% |
| XuanWu | 990459 | 2701024 | 63.3% |
| YuHuaTai | 497849 | 1387883 | 64.1% |
| QiXia | 917562 | 1787173 | 48.7% |
| PuKou | 752001 | 1521008 | 50.6% |
| LiuHe | 262370 | 690744 | 40.0% |
| JiangNing | 746350 | 1486711 | 49.8% |
| All | 7765393 | 16701097 | 53.5% |

The spatial distribution of administrative districts and pandemic control areas are shown in Figure 4. It can be seen that the boundary of the control areas does not match the administrative boundary. This means that the residents' scope of daily activities does not correspond to the administrative boundaries, so only controlling the COVID-19 pandemic through the administrative district will greatly lower the control effectiveness. However, the proposed model of zoning pandemic control areas based on the travel pattern of residents can effectively reduce the impact of COVID-19 on residents' travel. Juneau et al. (*34*) found that home quarantines are less cost-effective and social distancing measures like workplace and school closures are effective but costly, making them the least cost-effective options. On the premise of ensuring that necessary quarantine and social distancing measures in each zone, the method proposed in this paper reducing the impact of COVID-19 on residents' travel helps to reduce the cost of pandemic control, and achieve sustainable pandemic control. In addition, through Nanjing weekly multimodal transport ridership aggregated at the TAZ level and travel flows between TAZs, which are shown in Figure 5, for areas with high travel intensity and the regions with strong travel flows, the pandemic control areas can be used to manage key areas according to conditions of each partition. Strictly maintaining social distance in high-risk areas can reduce crowd gathering.

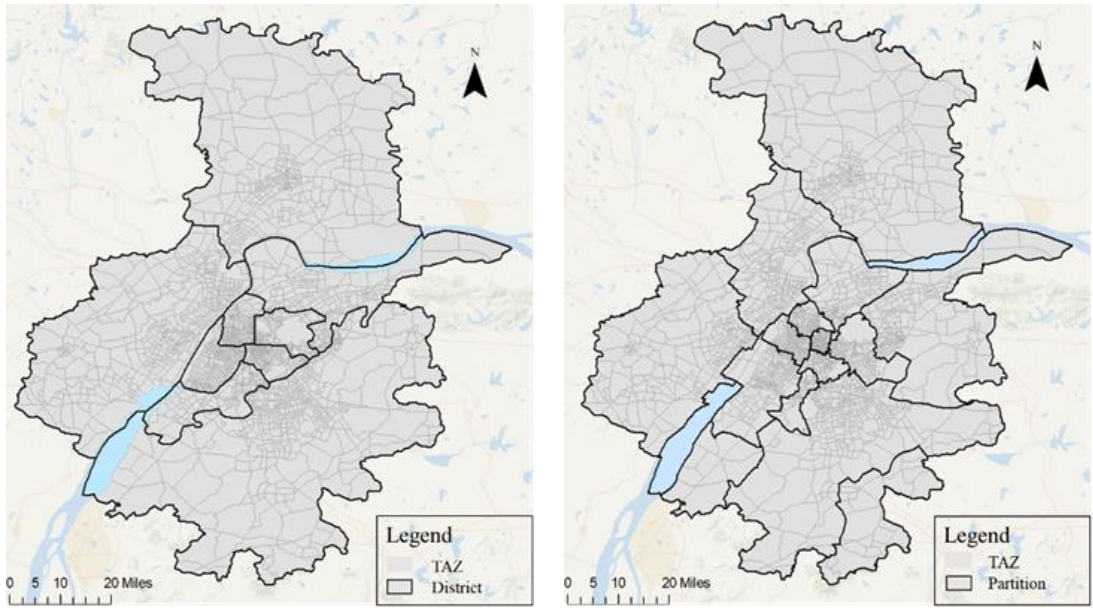

**(a)**                  **(b)**

**Figure 4 Spatial distribution of (a) administrative districts and (b) pandemic control areas**

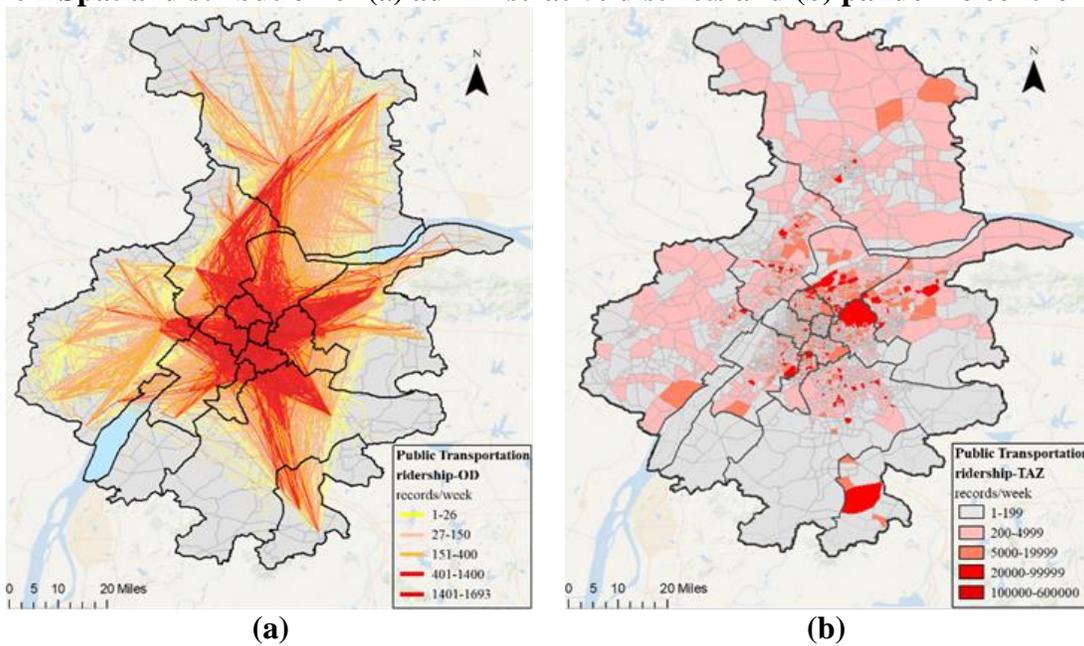

**(a)**                  **(b)**

**Figure 5 Nanjing weekly multimodal transport ridership: (a) aggregated at TAZ level and (b) travel flows between TAZs.**

        Due to the small granularity of the division of control zones in the city, the existing zoning results may not meet all the needs of pandemic control. We can change the level and scope of closed-off management based on the travel demands of residents and the spread of virus, which has the advantage of more sustainability and low cost under the background of a long-term situation of COVID-19. So, this paper proposes a strategy for merging control areas based on population, area, medical conditions, and difficulty in management and control, which helps to conduct rapid full COVID-19 screening and centralized management during the outbreak. Since

the trips between control areas are minimized, then after merging some partitions, the trips between them are still minimized. Figure 6 shows a merging result that tries to satisfy population balance and area balance, which divides Nanjing into three pandemic control areas. Among them, control area 1 mainly includes two administrative districts of Luhe and Pukou. Control area 2 mainly includes Jianye, Yuhuatai, and most areas of Jiangning. Control area 3 mainly includes Gulou, Qinhuai, Xuanwu, Qixia, and a few areas in Jiangning.

Table 5 shows some of the indicators of the three partitions after merging. It can be calculated that only 24.8% of travel connections are cut off, which reduces the impact of pandemic control on residents' travel. What's more, cutting off fewer travel connections is conducive to dealing with the long-term trend of the COVID-19.

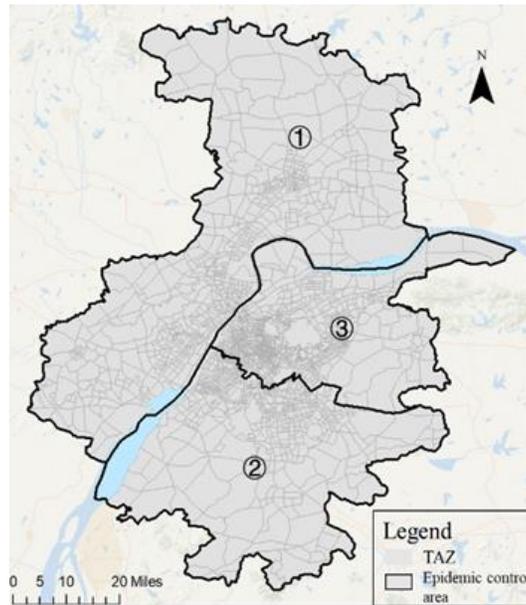

**Figure 6 Spatial distribution of partitions after merging**

**Table 5 Statistics of partitions after merging**

| ID | Area / km² | Population /10000 people | Travel within the partition | Total trips in partition | Percentage of cut-off trips |
|---|---|---|---|---|---|
| Partition① | 2420.41 | 240.13 | 1444050 | 2243413 | 35.6% |
| Partition② | 1517.95 | 319.90 | 2149669 | 3585166 | 40.0% |
| Partition③ | 884.97 | 513.63 | 8959600 | 10872518 | 17.6% |
| ALL | 4823.33 | 1073.66 | 12553319 | 16701097 | 24.8% |

## 6. DISCUSSION

Among the existing COVID-19 pandemic control measures, there are the following mainstream control measures according to the spread of the virus. When some local or overseas imported cases are found in the city, small-scale closed-off management measures are mainly carried out in the areas involved in their main activity scope, and personnel activities in local areas are restricted. When the COVID-19 virus spreads widely and more people in the city are diagnosed, there are two main modes of pandemic control. The first measure is to divide the pandemic control areas according to the administrative boundaries, and implement the closure management of administrative areas with severe epidemics. People in the closed-off areas are not allowed to travel across districts. For example, there is a large-scale outbreak at Beijing, China in 2022, and the Beijing government adopted the method of administrative control to respond to the COVID-19 pandemic. The second is to divide the pandemic control areas according to geographical interval, and implement regional control over the city through natural geographical interval. For example, at Shanghai, China in 2022, the city is divided into two control areas through the geographical interval of the Huangpu River for COVID-19 pandemic response. When the pandemic becomes uncontrollable, city-wide closed-off measures will be taken to cut off the spread of the pandemic across the country. For example, Wuhan 2020, China.

The effective control of the pandemic by the above measures is undeniable, but in the post-pandemic era, more flexible and refined pandemic response methods are needed. The pandemic zoning method based on multimodal urban transport data proposed in this paper is more flexible and refined for pandemic response in the post-pandemic era. First of all, the number of urban control zones can be flexibly set according to the pandemic control needs and pandemic control resources in the city. This paper gives two zoning plans. The first plan divides Nanjing into 15 pandemic control zones, which minimizes traffic flows between pandemic control zones. The second plan divides Nanjing into three control zones, and the balance of population and area between zones is considered. This pandemic zoning method redraws the pandemic control unit. Upgrade step-by-step control policy can be implemented through this pandemic zoning method. Based on zoning different levels of pandemic control areas, a gradual escalation of control policies to deal with the spread of the COVID-19 virus can be established. Specifically, the scope of pandemic control is gradually upgraded/downgraded according to the pandemic situation using outer edge risk assessment. When the first case appears, it is necessary to investigate its main movement trajectory, the main activity area needs to be locked down for pandemic control, and the travel of the residents in the area needs to be restricted. When the COVID-19 spreads further, the control scope of the current pandemic control unit cannot meet the sustainability requirements of long-term pandemic control, the surrounding control units can be merged to expand the area of pandemic control. Upgrading the level and scope of closed-off management should be based on the travel demands of residents and the spread of virus, which has the advantage of more sustainability and low cost under the background of a long-term situation of COVID-19. It provides valuable insights for realizing sustainable and refined pandemic control.

In addition, zoning pandemic screening and gradual reopening policy can be adopted based on the zoning results. Taking each pandemic control area as a unit can carry out rapid local COVID-19 screening. Each pandemic control unit adopts a rotating clearing reopen mode. That means, in all pandemic control areas in the city, gradually carry out nucleic acid testing for all employees, transfer infected persons, area disinfection, and decrease the number of infected people. In some areas, pandemic control will be carried out first, and all medical resources will be concentrated to achieve COVID-19 control in areas with the severe pandemic, and nucleic acid testing will be carried out first, and all asymptomatic or mild patients will be quarantined at home in other control areas. Once the number of infected people in the pandemic control area is cleared, work and living services in the area will resume simultaneously. And so on for other units for COVID-19 control until all positive infections are cleared in the city. Last but not least, multimodal transport operation and Internet Enterprise Operation Scope adjustment policy can be adopted according to the condition of pandemic zoning. The public transport network can be partitioned operated according to the zoning results. Take the pandemic control area as the operating scope of the multimodal transport network, and adjust the multimodal transport network within the scope. Multimodal transport will be suspended in the area where the COVID-19 occurs, while in other areas multimodal transport will operate within the area, thereby reducing the impact of the COVID-19 pandemic on travel within the city. Offline services of Internet companies, such as takeaway, are not allowed to dispatch orders across partitions and implement the policy that people cannot move across regions. All takeaway companies should make full use of technical means to achieve regional services and limit the movement of takeaway staff across the region. In addition, all the existing pandemic control strategies can be considered to combine with pandemic zoning control to achieve more scientific and effective pandemic control.

Different from administrative divisions and geographic intervals, pandemic control partitions based on traffic travel data can better reflect people's travel demands and reduce the impact of pandemic control on residents' lives and work. To a certain extent, this method reduces the consumption of pandemic response and contributes to sustainable pandemic control in the post-epidemic era. From the statistics of the results, in the zoning scheme of the 15 pandemic control zones, the total percentage of cut-off trips is 45.3%, and the proportion of trips cut off in each partition is counted separately, and no high cut-off ratio appears. Compared with the zoning scheme by administrative division, which has a total percentage of cut-off trips of 53.5%. The scheme in this paper is reduced by 8.2%, and it is worth noting that the number of partitions according to the administrative area is 9 categories and less than 15 division schemes. Obviously, the more the number of partitions, the higher the cut-off trips ratio, but the plan in this paper is still lower than the administrative district plan. In addition, the statistical results of the three big pandemic control zones are also excellent, and the total cut-off trips ratio is 24.8%. It can be seen that the method is significant in reducing the impact of pandemic control on residents' travel.

## 7. CONCLUSIONS

In the post-pandemic era, the low-cost and sustainable fight against COVID-19 has become more and more significant. The difficulty of COVID-19 control is closely related to the size of the population and the area of the control regions. Scientific partition of reasonable pandemic control areas to reduce the impact of COVID-19 on residents' daily life has become an urgent problem to be solved. Partitioning of pandemic control areas through community detection methods can effectively reduce the difficulty of pandemic control. It can maximize travel within the pandemic control areas and minimize travel connections between pandemic control areas, which can reduce the impact of COVID-19 on residents' daily life and work, and control the spread of COVID-19 effectively. Thus, reducing the cost of pandemic control and realizing refined pandemic control within the city. Therefore, this paper aims to scientifically zone the urban pandemic control areas by using the community detection algorithm based on Nanjing multimodal transport data.

First, this paper builds a spatial interaction network based on the multimodal transport travel data in Nanjing. Second, based on the requirements of pandemic control, the Leiden algorithm improved by the gravity model is used to partition the pandemic control areas. Then, communities with weak spatial connectivity are corrected by adding three constraints. Finally, 15 small pandemic control areas and 3 big pandemic control areas considering geospatial relationships were acquired. Different from the administrative area, the control areas take into account the spatial connection of residents' travel. Finally, this paper proposes recommendations for COVID-19 control according to the results of pandemic zoning and travel characteristics in the region. This paper is expected to contribute to reducing the cost of pandemic control in the post-pandemic era and achieving sustainable and refined pandemic response.

Zoning pandemic control areas meet the needs of citizens and are of full significance to refined and sustainable COVID-19 control. At the same time, it provides insights for control policies during COVID-19. Although the zoning management model breaks the historical evolution of administrative district management, this initiative is feasible under the unified leadership of the government, as well as the cooperation between partitions. However, in this paper, only two-division schemes in Nanjing are given, and the partition plans of other intermediate states are not given. Therefore, the division scheme can be adjusted according to the city's situation, such as the pandemic risk level, the city's pandemic' prevention resources, the city's population, and area, etc. These factors will have an impact on the results of pandemic control zoning. Therefore, it is necessary to determine a reasonable number of control areas according to local conditions to realize the promotion of the method of zoning the pandemic control areas in the city.

Admittedly, this study comes with some limitations, which should be addressed in future studies. First, the Leiden algorithm needs to be compared with other community detection algorithms and clustering algorithms to improve the effectiveness of pandemic control partitioning in the future. Second, multimodal transport data can only reflect part of the characteristics of urban residents' travel. Mobile phone signaling data can be used to conduct related research, which may make the partition results more accurate and reasonable. Third, the balance of population and area between pandemic control areas are not considered in this method, which may lead to inconsistencies in the difficulty of pandemic control in various regions. A balance between population and area in community detection results can be achieved in the future.

**ACKNOWLEDGMENTS**

**AUTHOR CONTRIBUTIONS**

The authors confirm their contribution to the paper as follows: Yufei Wang: Study conception and Design, Data Preprocessing, Analysis and Interpretation of Results, Draft Manuscript Preparation. Mingzhuang Hua: Study conception and Design, Analysis and Interpretation of Results. Xuewu Chen: Study conception and Design, Reviewing and Editing. Wendong Chen: Data Preprocessing, Visualization. Long Cheng: Reviewing and Editing. All authors reviewed the results and approved the final version of the manuscript.